\documentclass[journal]{new-aiaa}

\usepackage[utf8]{inputenc}
\usepackage{textcomp}

\usepackage{graphicx}
\usepackage{amsmath}
\usepackage[version=4]{mhchem}
\usepackage{siunitx}
\usepackage{longtable,tabularx}
\title{Closed-Form Expressions for Five-Digit Reflex Camber-Line Design Parameters}

\author{Kio M. Lovric\footnote{Undergraduate Researcher, School of Aerospace, Mechanical and Mechatronic Engineering and School of Physics; Student Member AIAA.}}
\affil{The University of Sydney, Sydney, NSW 2006, Australia}

\begin{document}

\maketitle

\begin{abstract}
Despite nearly a century of use in tailless and flying-wing aircraft, the NACA five-digit reflex camber-line family lacks published closed-form expressions for the governing design integrals; practitioners have instead relied on numerical quadrature and tabulated constants available only for a limited set of standard configurations. This paper addresses that gap by deriving closed-form analytical expressions for all lift and zero-moment integrals in terms of elementary functions of the breakpoint and maximum-camber-location parameters. A trigonometric substitution eliminates the endpoint singularities introduced by the Glauert transformation, yielding integrals expressible as inverse trigonometric functions combined with explicit polynomials. The breakpoint parameter is determined by solving a single transcendental equation, and the remaining camber-line constants follow from direct evaluation without numerical integration. Independent verification by numerical quadrature confirms agreement with the analytical expressions to machine precision. Comparison with historical tabulations shows that the closed-form values satisfy the zero-moment design condition to machine precision, whereas substituting the tabulated breakpoints back into the same condition yields residuals nine to thirteen orders of magnitude larger, consistent with the rounding and computational precision available when the original tabulations were produced.
\end{abstract}

\section*{Nomenclature}

{\renewcommand\arraystretch{1.0}%
\noindent\begin{longtable*}{@{}l @{\quad=\quad} l@{}}
$a_0, a_1, a_2, a_3, a_4$ & polynomial coefficients in forward-region integral derivations \\
$A, B, C_3, C_5, C_7$     & grouped integration coefficients for forward region \\
$A', B', C'_3, C'_5, C'_7$ & grouped integration coefficients for aft region \\
$A_0, A_1, A_2$           & Fourier coefficients of camber-line slope \\
$A_n$                     & $n$th Fourier coefficient of camber-line slope \\
$b_0, b_1, b_2$           & polynomial coefficients in aft-region slope expression \\
$c$                       & chord, m \\
$c_0, c_1, c_2, c_3, c_4$ & polynomial coefficients in aft-region integral derivations \\
$C_l$                     & section lift coefficient \\
$C_{l,i}$                 & design lift coefficient \\
$C_{m_{c/4}}$             & pitching-moment coefficient about quarter-chord \\
$\mathcal{D}$             & denominator polynomial in $k_1$ expression \\
$\mathrm{d}y_c/\mathrm{d}x$ & camber-line slope \\
$I_1^{(l)}, I_2^{(l)}$    & lift-condition integrals \\
$I_1^{(m)}, I_2^{(m)}$    & moment-condition integrals \\
$k_1$                     & primary camber-line scaling constant \\
$k_2$                     & secondary camber-line scaling constant \\
$L$                       & first digit of NACA designation ($C_{l,i} = 0.15L$) \\
$n$                       & Fourier series index \\
$P$                       & second digit of NACA designation ($x_{mc} = P/20$) \\
$\mathcal{P}$             & polynomial in $I_1^{(m)}$ expression \\
$Q$                       & third digit of NACA designation (0 = standard, 1 = reflex) \\
$\mathcal{Q}$             & polynomial in $I_2^{(m)}$ expression \\
$r$                       & camber-line breakpoint location \\
$\mathcal{R}$             & polynomial in $I_2^{(m)}$ expression \\
$\mathcal{S}$             & polynomial in $I_1^{(l)}$ expression \\
$t$                       & airfoil thickness, m \\
$\mathcal{T}$             & polynomial in $I_2^{(l)}$ expression \\
$TT$                      & thickness digits ($t/c = TT/100$) \\
$\mathcal{U}$             & polynomial in $I_2^{(l)}$ expression \\
$x$                       & chordwise coordinate \\
$x_{mc}$                  & maximum camber location \\
$y_c$                     & camber-line ordinate \\
\multicolumn{2}{@{}l}{}\\
\multicolumn{2}{@{}l}{\textit{Greek Symbols}}\\
$\alpha$                  & angle of attack, rad \\
$\alpha_i$                & ideal angle of attack, rad \\
$\theta$                  & Glauert transformation variable, rad \\
$\phi$                    & integration substitution variable, rad \\
$\phi_r$                  & value of $\phi$ at breakpoint, $\arcsin(\sqrt{r})$, rad \\
$\psi_r$                  & complementary angle, $\pi/2 - \phi_r$, rad \\
\multicolumn{2}{@{}l}{}\\
\multicolumn{2}{@{}l}{\textit{Subscripts}}\\
anal                      & analytical evaluation \\
$c/4$                     & quarter-chord \\
$i$                       & ideal (design) condition \\
mc                        & maximum camber \\
quad                      & numerical quadrature \\
ref                       & tabulated reference \\
\multicolumn{2}{@{}l}{}\\
\multicolumn{2}{@{}l}{\textit{Superscripts}}\\
$(l)$                     & lift condition \\
$(m)$                     & moment condition \\
\end{longtable*}}

\section{Introduction}

\lettrine{T}{ailless} and flying-wing configurations often employ reflexed camber lines to achieve near-zero pitching moment about the quarter-chord, thereby reducing reliance on a horizontal stabilizer for longitudinal trim~\cite{Abbott_Doenhoff}. Despite decades of use dating to the 1930s~\cite{NACA_5digit_1935}, determination of the breakpoint location $r$ and the scaling constant $k_1$ has traditionally relied on numerical evaluation and tabulated values, which exist only for a limited set of standard configurations~\cite{Ladson_Brooks_1975}.

In the standard NACA designation $LPQTT$, the first digit $L$ sets the design lift coefficient $C_{l,i}=0.15\, L$, the second digit $P$ locates the maximum camber at $x_{mc}=P/20$, the third digit $Q$ selects the camber-line family ($Q=0$ for standard, $Q=1$ for reflex), and the final two digits $TT$ specify the thickness ratio $t/c=TT/100$~\cite{Abbott_Doenhoff}. For the reflex family, $r$ is chosen to enforce the zero-moment condition $C_{m_{c/4}}=0$, and $k_1$ is then set to satisfy the prescribed design lift coefficient.

While thin-airfoil theory provides closed-form coefficient relations for general camber lines~\cite{Glauert_TAT}, the specific integrals governing the reflex design conditions do not appear to be published in closed form in the standard NACA/NASA tabulations. Here, we derive closed-form expressions for all governing integrals, namely the moment integrals $I_1^{(m)}$ and $I_2^{(m)}$ and the lift integrals $I_1^{(l)}$ and $I_2^{(l)}$, in terms of elementary functions of $r$ and $x_{mc}$. The resulting expressions eliminate numerical quadrature: $r$ is found by solving $I_1^{(m)}+\frac{1}{(1-r)^3}I_2^{(m)}=0$ with a single-variable root-finder, and $k_1$ then follows from direct evaluation.

We validate the analytical results against NASA tabulated values~\cite{Ladson_Brooks_1975} and independent numerical quadrature. The remainder of this paper is organized as follows: Section~\ref{sec:math_formulation} develops the mathematical formulation; Section~\ref{sec:closed_form_results} presents the closed-form results; Section~\ref{sec:design_procedure} describes the design procedure; Section~\ref{sec:validation} presents the validation, and Section~\ref{sec:extended_parameters} provides tabled and graphical extended design parameters. Detailed derivations are provided in Appendices~\ref{sec:moment_derivation}--\ref{app:lift_derivation}.

\section{Mathematical Formulation}\label{sec:math_formulation}

\subsection{Reflex Camber-Line Geometry}
The NACA 5-digit reflex camber line is defined piecewise with maximum camber location parameter $x_{mc}=P/20$, break location $r$, and scaling constants $k_1$ and $k_2$. The camber-line ordinate is
\begin{equation}
y_c = \begin{cases}
\dfrac{k_1}{6}\left[(x-r)^3 - \dfrac{k_2}{k_1}(1-r)^3 x - r^3 x + r^3\right], & 0 \leq x < r \\
\dfrac{k_1}{6}\left[\dfrac{k_2}{k_1}(x-r)^3 - \dfrac{k_2}{k_1}(1-r)^3 x - r^3 x + r^3\right], & r \leq x \leq 1
\end{cases}
\label{eq:yc}
\end{equation}

The corresponding camber-line slope is
\begin{equation}
\frac{\mathrm{d}y_c}{\mathrm{d}x}
 = \begin{cases}
\dfrac{k_1}{6}\left[3(x-r)^2 - \dfrac{k_2}{k_1}(1-r)^3 - r^3\right], & 0 \leq x < r \\
\dfrac{k_1}{6}\left[3\dfrac{k_2}{k_1}(x-r)^2 - \dfrac{k_2}{k_1}(1-r)^3 - r^3\right], & r \leq x \leq 1
\end{cases}
\label{eq:slope}
\end{equation}

The ratio $k_2/k_1$ is determined by requiring maximum camber to occur at $x = x_{mc}$, giving
\begin{equation}
\frac{k_2}{k_1} = \frac{3(r - x_{mc})^2 - r^3}{(1-r)^3}
\label{eq:k2k1_ratio}
\end{equation}
Substituting Eq.~\eqref{eq:k2k1_ratio} into Eq.~\eqref{eq:slope} and simplifying yields the working forms of the camber-line slope, written separately for each region. For $0 \leq x < r$,
\begin{equation}
\left(\frac{\mathrm{d}y_c}{\mathrm{d}x}\right)_{<r} = \frac{k_1}{2}\left(x^2 - 2rx + 2x_{mc}r - x_{mc}^2\right)
\label{eq:slope_forward_simplified}
\end{equation}
and for $r \leq x \leq 1$,
\begin{equation}
\left(\frac{\mathrm{d}y_c}{\mathrm{d}x}\right)_{\geq r} = \frac{k_1}{2(1-r)^3}\left(b_2 x^2 + b_1 x + b_0\right)
\label{eq:slope_aft_simplified}
\end{equation}
where the polynomial coefficients are
\begin{align}
b_2 &= 3(r - x_{mc})^2 - r^3 \label{eq:b2}\\
b_1 &= 2r^4 - 6r^3 + 12r^2 x_{mc} - 6r x_{mc}^2 \label{eq:b1}\\
b_0 &= 3r^3 - r^2 - 2r^4 x_{mc} - 6r^2 x_{mc} + 2r x_{mc} + r^3 x_{mc}^2 + 3r x_{mc}^2 - x_{mc}^2 \label{eq:b0}
\end{align}

These camber-line slope expressions provide the integrands required by thin-airfoil theory to determine the aerodynamic coefficients.

\subsection{Thin Airfoil Theory}
In classical thin-airfoil theory~\cite{Glauert_TAT,Abbott_Doenhoff}, the camber-line slope is expanded in a Fourier cosine series using the Glauert transformation
\begin{equation}
x = \frac{1 - \cos\theta}{2}, \quad \theta \in [0, \pi]
\label{eq:glauert_transform}
\end{equation}

The Fourier coefficients are
\begin{align}
A_0 &= \alpha - \frac{1}{\pi}\int_0^{\pi} \frac{\mathrm{d}y_c}{\mathrm{d}x} \, \mathrm{d}\theta \label{eq:A0}\\
A_n &= \frac{2}{\pi}\int_0^{\pi} \frac{\mathrm{d}y_c}{\mathrm{d}x} \cos(n\theta) \, \mathrm{d}\theta, \quad  n=1,2,\ldots \label{eq:An}
\end{align}

The section lift and moment coefficients follow as
\begin{align}
C_l &= 2\pi\left(A_0 + \frac{A_1}{2}\right) \label{eq:Cl}\\
C_{m_{c/4}} &= \frac{\pi}{4}(A_2 - A_1) \label{eq:cm_quarter_chord}
\end{align}

\subsection{Design Constraints}

\subsubsection{Constraint 1: Zero Pitching Moment}
For a reflex camber line, the primary design requirement is zero pitching moment about the quarter-chord such that
\begin{equation}
C_{m_{c/4}} = 0
\label{eq:zero_moment}
\end{equation}
From Eq.~\eqref{eq:cm_quarter_chord}, this requires $A_2 - A_1 = 0$, which leads to
\begin{equation}
\int_0^{\pi} \frac{\mathrm{d}y_c}{\mathrm{d}x} \left(\cos 2\theta - \cos\theta\right) \mathrm{d}\theta = 0
\label{eq:zero_moment_integral}
\end{equation}
This constraint is homogeneous in $k_1$ and determines $r$ (and hence $k_2/k_1$) by selecting the physically admissible root $r \in (x_{mc},1)$; however, since $k_1$ cancels from the equation, the absolute camber magnitude is not fixed by this condition alone.

\subsubsection{Constraint 2: Design Lift Coefficient}
The ideal angle of attack $\alpha_i$ is defined by $A_0(\alpha_i)=0$, which eliminates the leading-edge singularity. The design lift coefficient is then
\begin{equation}
C_{l,i} = \pi A_1 = 2\int_0^{\pi} \frac{\mathrm{d}y_c}{\mathrm{d}x} \cos\theta \, \mathrm{d}\theta
\label{eq:Cli_definition}
\end{equation}
This constraint, together with the prescribed design lift coefficient, uniquely determines the scaling constant $k_1$.

\subsection{Transformation to \texorpdfstring{$x$}{x}-Domain}
From the Glauert mapping in Eq.~\eqref{eq:glauert_transform}, the inverse relations are
\begin{align}
\cos\theta &= 1-2x \label{eq:cos_theta_x}\\
\sin\theta &= 2\sqrt{x(1-x)} \label{eq:sin_theta_x}
\end{align}
For the moment condition,
\begin{equation}
\cos 2\theta - \cos\theta
= \left(2\cos^2\theta-1\right)-\cos\theta
= 8x^2-6x
\label{eq:cos_transform}
\end{equation}
while for the lift condition, $\cos\theta = 1-2x$ follows directly from Eq.~\eqref{eq:cos_theta_x}. The Jacobian associated with the mapping is
\begin{equation}
\mathrm{d}\theta=\frac{\mathrm{d}x}{\sqrt{x(1-x)}}
\label{eq:jacobian}
\end{equation}
so that $\theta\in[0,\pi]$ corresponds to $x\in[0,1]$.

\subsection{Formulation of the Design Integrals}
Both design constraints reduce to weighted integrals over $x\in[0,1]$ with the common Jacobian
${\mathrm{d}x}/\sqrt{x(1-x)}$ arising from the Glauert mapping. Because the reflex camber-line slope is piecewise-defined with a breakpoint at $x=r$, each constraint naturally splits into a forward contribution
($0\le x<r$) and an aft contribution ($r\le x\le 1$). The resulting integrals share the same
integrable endpoint singularities associated with $1/\sqrt{x(1-x)}$; these are removed in the
closed-form evaluation by the substitution $x=\sin^2\phi$ (see Section~\ref{sec:trig_sub}).

\subsubsection{Moment Condition Integrals}
Substituting Eq.~\eqref{eq:jacobian} into Eq.~\eqref{eq:zero_moment_integral} and using
Eq.~\eqref{eq:cos_transform}, the zero-moment condition becomes
\begin{equation}
\int_0^{1} \frac{\mathrm{d}y_c}{\mathrm{d}x}\frac{8x^2-6x}{\sqrt{x(1-x)}}\,{\mathrm{d}x} = 0
\label{eq:moment_xdomain}
\end{equation}
Splitting at $x=r$ and substituting the simplified slope expressions
Eqs.~\eqref{eq:slope_forward_simplified} and~\eqref{eq:slope_aft_simplified} yields
\begin{equation}
\frac{k_1}{2}\left[
I_1^{(m)} + \frac{1}{(1-r)^3}I_2^{(m)}
\right]=0
\label{eq:moment_split_reduced}
\end{equation}
where the reduced moment integrals, independent of $k_1$, are defined as
\begin{align}
I_1^{(m)} &\equiv \int_0^{r}
\left(x^2-2rx+2x_{mc}r-x_{mc}^2\right)\frac{8x^2-6x}{\sqrt{x(1-x)}}\,{\mathrm{d}x}
\label{eq:I1m_def}\\
I_2^{(m)} &\equiv \int_r^{1}
\left(b_2x^2+b_1x+b_0\right)\frac{8x^2-6x}{\sqrt{x(1-x)}}\,{\mathrm{d}x}
\label{eq:I2m_def}
\end{align}
Since $k_1 \neq 0$ for a nontrivial camber line, it may be divided out, and the breakpoint parameter $r$ is determined by
\begin{equation}
I_1^{(m)} + \frac{1}{(1-r)^3}I_2^{(m)} = 0
\label{eq:design_condition_moment}
\end{equation}

\subsubsection{Lift Condition Integrals}
The lift constraint follows in the same manner, with $\cos\theta=1-2x$.
Substituting Eq.~\eqref{eq:jacobian} into Eq.~\eqref{eq:Cli_definition} gives
\begin{equation}
C_{l,i} = 2\int_0^{1} \frac{\mathrm{d}y_c}{\mathrm{d}x}\,\frac{1-2x}{\sqrt{x(1-x)}}\,{\mathrm{d}x}
\label{eq:Cli_transformed}
\end{equation}
Splitting at $x=r$ and substituting Eqs.~\eqref{eq:slope_forward_simplified}
and~\eqref{eq:slope_aft_simplified} yields
\begin{equation}
C_{l,i} = k_1\left[I_1^{(l)} + \frac{1}{(1-r)^3}I_2^{(l)}\right]
\label{eq:Cli_integrals}
\end{equation}
where
\begin{align}
I_1^{(l)} &\equiv \int_0^{r}
\left(x^2-2rx+2x_{mc}r-x_{mc}^2\right)\frac{1-2x}{\sqrt{x(1-x)}}\,{\mathrm{d}x}
\label{eq:I1l_def}\\
I_2^{(l)} &\equiv \int_r^{1}
\left(b_2x^2+b_1x+b_0\right)\frac{1-2x}{\sqrt{x(1-x)}}\,{\mathrm{d}x}
\label{eq:I2l_def}
\end{align}
The prescribed design lift coefficient $C_{l,i}=0.15L$ then determines the scaling constant
\begin{equation}
k_1 = \frac{C_{l,i}}{I_1^{(l)} + \dfrac{1}{(1-r)^3}I_2^{(l)}}
\label{eq:k1_formula}
\end{equation}

\subsection{Trigonometric Substitution}\label{sec:trig_sub}
The design integrals in Eqs.~\eqref{eq:I1m_def}--\eqref{eq:I2l_def} contain the Glauert Jacobian
$1/\sqrt{x(1-x)}$, which produces integrable algebraic endpoint singularities at $x=0$ and $x=1$
(typically $\sim x^{-1/2}$ as $x\to 0$ and $\sim (1-x)^{-1/2}$ as $x\to 1$, up to the polynomial
numerators). Although the integrals converge, these square-root factors obscure direct antiderivative
evaluation in the $x$-domain. A standard remedy is to introduce an angle variable for which the
Jacobian cancels identically.

Accordingly, we apply the substitution
\begin{equation}
x=\sin^2\phi,
\qquad
{\mathrm{d}x} = 2\sin\phi\cos\phi\,{\mathrm{d}\phi},
\qquad
\sqrt{x(1-x)}=\sin\phi\cos\phi
\label{eq:trig_sub}
\end{equation}
so that the weighted measure simplifies to a constant,
\begin{equation}
\frac{{\mathrm{d}x}}{\sqrt{x(1-x)}} = 2\,{\mathrm{d}\phi}
\label{eq:phi_measure}
\end{equation}
Under this mapping, the integrands become polynomial combinations of $\sin^{2n}\phi$ (and, where
required, $\cos^{2n}\phi$) with constant measure ${\mathrm{d}\phi}$, enabling closed-form evaluation via
standard trigonometric antiderivatives. The limits map as
\begin{equation}
x=0 \Rightarrow \phi=0,
\qquad
x=r \Rightarrow \phi=\phi_r \equiv \arcsin(\sqrt{r}),
\qquad
x=1 \Rightarrow \phi=\frac{\pi}{2}
\end{equation}
It is often convenient to note the complementary relation
$\arccos(\sqrt{r})=\frac{\pi}{2}-\phi_r$ for the aft-region expressions.

\section{Closed-Form Results}\label{sec:closed_form_results}

For clarity, we present the closed-form \emph{definite} expressions first; full derivations appear in
Appendices~\ref{sec:moment_derivation}--\ref{app:lift_derivation}. 

\subsection{Moment Condition Integrals}

\subsubsection{Integral \texorpdfstring{$I_1^{(m)}$}{I1m} over \texorpdfstring{$0 \leq x < r$}{0 <= x < r}}
\begin{equation}
I_1^{(m)} =
\frac{5-8r}{8}\,\arcsin(\sqrt{r})
+\sqrt{r(1-r)}\,\mathcal{P}(r,x_{mc})
\label{eq:I1m_closed}
\end{equation}
where
\begin{equation}
\mathcal{P}(r,x_{mc})
=4r\big(x_{mc}-r\big)^2
+\frac{-16r^3+8r^2+14r-15}{24}
\label{eq:P_polynomial}
\end{equation}

\subsubsection{Integral \texorpdfstring{$I_2^{(m)}$}{I2m} over \texorpdfstring{$r \leq x \leq 1$}{r <= x <= 1}}
\begin{equation}
I_2^{(m)} =
\mathcal{Q}(r,x_{mc})\,\arccos(\sqrt{r})
-\sqrt{r(1-r)}\,\mathcal{R}(r,x_{mc})
\label{eq:I2m_closed}
\end{equation}
where
\begin{equation}
\mathcal{Q}(r,x_{mc})
=\frac{8r-5}{8}\left[r^3-3\big(x_{mc}-r\big)^2\right]
\label{eq:Q_polynomial}
\end{equation}
and
\begin{equation}
\mathcal{R}(r,x_{mc})
=\frac{r^3}{24}\big(16r^3-8r^2-14r+15\big)
-\frac{1}{8}\big(32r^4-80r^3+88r^2-46r+15\big)\big(x_{mc}-r\big)^2
\label{eq:R_polynomial}
\end{equation}

\subsection{Lift Condition Integrals}

\subsubsection{Integral \texorpdfstring{$I_1^{(l)}$}{I1l} over \texorpdfstring{$0 \leq x < r$}{0 <= x < r}}
\begin{equation}
I_1^{(l)} =
\left(r-\frac{1}{2}\right)\arcsin(\sqrt{r})
+\sqrt{r(1-r)}\,\mathcal{S}(r,x_{mc})
\label{eq:I1l_closed}
\end{equation}
where
\begin{equation}
\mathcal{S}(r,x_{mc})
=\frac{1}{3}+\frac{2}{3}\left(r-\frac{1}{2}\right)^2-2\big(x_{mc}-r\big)^2
\label{eq:S_polynomial}
\end{equation}

\subsubsection{Integral \texorpdfstring{$I_2^{(l)}$}{I2l} over \texorpdfstring{$r \leq x \leq 1$}{r <= x <= 1}}
\begin{equation}
I_2^{(l)} =
\mathcal{T}(r,x_{mc})\,\arccos(\sqrt{r})
-\sqrt{r(1-r)}\,\mathcal{U}(r,x_{mc})
\label{eq:I2l_closed}
\end{equation}
where
\begin{equation}
\mathcal{T}(r,x_{mc})
=\left(\frac{1}{2}-r\right)\left[r^3-3\big(x_{mc}-r\big)^2\right]
\label{eq:T_polynomial}
\end{equation}
and
\begin{equation}
\mathcal{U}(r,x_{mc})
=\frac{1}{2}\big(4r^3-8r^2+8r-1\big)\big(x_{mc}-r\big)^2
-\frac{r^3}{6}\big(4r^2-4r+3\big)
\label{eq:U_polynomial}
\end{equation}

\section{Design Procedure}\label{sec:design_procedure}

\subsection{Solution for Breakpoint Parameter \texorpdfstring{$r$}{r}}

The zero-pitching-moment condition requires $I_1^{(m)} + \frac{1}{(1-r)^3}I_2^{(m)} = 0$. Substituting the closed-form expressions yields a transcendental equation in $r$ and $x_{mc}$ only:
\begin{equation}
\begin{aligned}
\frac{5-8r}{8}\,\arcsin(\sqrt{r})
&+ \sqrt{r(1-r)}\,\mathcal{P}(r, x_{mc}) \\
&+ \frac{\mathcal{Q}(r, x_{mc})\,\arccos(\sqrt{r})
- \sqrt{r(1-r)}\,\mathcal{R}(r, x_{mc})}{(1-r)^3} = 0
\end{aligned}
\label{eq:r_transcendental}
\end{equation}

This equation is solved numerically for $r$ given $x_{mc}$, but crucially, no numerical integration is required; all terms are evaluated in closed form. When multiple roots exist in $(0,1)$, the breakpoint must satisfy $r > x_{mc}$ so that maximum camber occurs forward of the reflex region; the smallest such root is selected. For the standard range $P \in [1,5]$, the transcendental equation admits a unique root in $(x_{mc}, 1)$, and the left-hand side of Eq.~\eqref{eq:r_transcendental} is monotonic in this interval, ensuring robust convergence for standard root-finding algorithms (see Fig.~\ref{fig:validation_r}, lower-right panel). Equivalently, the function $f(r; x_{mc}) := I_1^{(m)} + (1-r)^{-3} I_2^{(m)}$ has exactly one zero on $(x_{mc}, 1)$ for $P \in [1,5]$, so bracketing in this interval guarantees convergence. As $P$ increases beyond 5, the root approaches unity and the reflex curvature becomes impractically large; for $P = 9$, Eq.~\eqref{eq:r_transcendental} admits no admissible root $r\in(x_{mc},1)$. Physically, the breakpoint $r$ represents the chordwise location where the camber-line curvature changes sign; moving maximum camber forward (decreasing $x_{mc}$) requires a correspondingly forward breakpoint to maintain the moment balance needed for zero pitching moment.

\subsection{Solution for Scaling Constant \texorpdfstring{$k_1$}{k1}}

Once $r$ is determined, the scaling constant $k_1$ follows directly from Eq.~\eqref{eq:k1_formula}. Substituting the closed-form expressions~\eqref{eq:I1l_closed}--\eqref{eq:I2l_closed}:
\begin{equation}
k_1 = \frac{C_{l,i}}{\mathcal{D}(r, x_{mc})}
\label{eq:k1_closed_form}
\end{equation}
where the denominator is
\begin{equation}
\begin{aligned}
\mathcal{D}(r, x_{mc}) &= \left(r - \frac{1}{2}\right)\arcsin(\sqrt{r}) + \sqrt{r(1-r)}\,\mathcal{S}(r, x_{mc}) \\
&\quad + \frac{1}{(1-r)^3}\left[\mathcal{T}(r, x_{mc})\,\arccos(\sqrt{r}) - \sqrt{r(1-r)}\,\mathcal{U}(r, x_{mc})\right]
\end{aligned}
\label{eq:D_denominator}
\end{equation}

\subsection{Complete Design Algorithm}
The preceding results can be assembled into a straightforward procedure for computing the camber-line parameters directly from a NACA designation. Given a NACA $LPQTT$ designation with $Q=1$ (reflex):
\begin{enumerate}
\item \textit{Input:} Extract $L$ and $P$ from the designation. Compute $x_{mc} = P/20$ and $C_{l,i} = 0.15L$.
\item \textit{Solve for $r$:} Find the root of Eq.~\eqref{eq:r_transcendental} in the interval $(x_{mc}, 1)$ using a one-dimensional root-finding algorithm (e.g., bisection, secant, Newton--Raphson, Brent's method, or regula falsi).
\item \textit{Compute $k_2/k_1$:} Evaluate Eq.~\eqref{eq:k2k1_ratio}.
\item \textit{Compute $k_1$:} Evaluate Eq.~\eqref{eq:k1_closed_form} with the known values of $r$, $x_{mc}$, and $C_{l,i}$.
\item \textit{Compute $k_2$:} $k_2 = (k_2/k_1) \cdot k_1$.
\item \textit{Output:} Complete camber-line parameters $(r, k_1, k_2)$.
\end{enumerate}
All steps involve only closed-form evaluations and a single root-finding operation; no numerical quadrature is required. When $r$ approaches $x_{mc}$, the ratio $k_2/k_1$ becomes sensitive to rounding, so double-precision arithmetic is recommended throughout. The expressions involving $\arcsin(\sqrt{r})$ and $\arccos(\sqrt{r})$ are numerically well-conditioned for all $r \in (0,1)$, and no cancellation issues arise in the polynomial terms.

\section{Validation and Discussion}\label{sec:validation}

The closed-form expressions derived in the preceding sections were validated against tabulated reference values for the NACA 5-digit reflex camber-line series as compiled by Ladson and Brooks~\cite{Ladson_Brooks_1975}. The analytically computed breakpoint parameter $r$ and scaling constants $k_1$ and $k_2/k_1$ are compared with tabulated values, and the moment and lift integrals are independently verified by numerical quadrature of the original definitions. A sensitivity analysis is then performed to determine whether observed discrepancies arise from the propagation of small differences in $r$ or from other sources. Graphical comparisons supplement the tabulated results.

\subsection{Validation of Breakpoint Parameter \texorpdfstring{$r$}{r}}
\label{subsec:validation_r}
The design parameter $r$ was obtained by solving the design constraint
\begin{equation}
I_1^{(m)}(r;x_{mc}) + \frac{1}{(1-r)^3}I_2^{(m)}(r;x_{mc}) = 0
\end{equation}
The root was computed using MATLAB's \texttt{fzero}, with $I_1^{(m)}$ and $I_2^{(m)}$ evaluated analytically using the derived closed forms. Table~\ref{tab:validation_r} compares the resulting values with the tabulations~\cite{Ladson_Brooks_1975}.

\begin{table}[hbt!]
\caption{\label{tab:validation_r} Closed-form breakpoint parameter $r$ compared with tabulated values~\cite{Ladson_Brooks_1975}.}
\centering
\begin{tabular}{lccccc}
\hline
Designation & $x_{mc}$ & $r_{\mathrm{ref}}$ & $r_{\mathrm{anal}}$ & $r_{\mathrm{anal}}-r_{\mathrm{ref}}$ & $\left|\left(I_1^{(m)}+\frac{1}{(1-r)^3}I_2^{(m)}\right)_{\mathrm{anal}}\right|$ \\
\hline
221 & 0.10 & 0.1300 & 0.1307 & $+7.50\times 10^{-4}$ & $3.59\times 10^{-17}$ \\
231 & 0.15 & 0.2170 & 0.2160 & $-9.85\times 10^{-4}$ & $2.78\times 10^{-15}$ \\
241 & 0.20 & 0.3180 & 0.3179 & $-8.10\times 10^{-5}$ & $1.09\times 10^{-15}$ \\
251 & 0.25 & 0.4410 & 0.4408 & $-1.70\times 10^{-4}$ & $3.99\times 10^{-17}$ \\
\hline
\end{tabular}
\end{table}

The maximum absolute discrepancy is $9.85\times 10^{-4}$, with a mean of $4.97\times 10^{-4}$. The residuals $|(I_1^{(m)}+\frac{1}{(1-r)^3}I_2^{(m)})_{\mathrm{anal}}|$ at the analytical solutions are $10^{-15}$--$10^{-17}$, indicating near machine-precision satisfaction of the design constraint. For reproducibility, the computed values to eight decimal places are provided in Appendix~\ref{app:validation_precision}.

\subsection{Independent Quadrature Verification of Moment Integrals}
\label{subsec:validation_quadrature}
To independently verify the analytical expressions for $I_1^{(m)}$ and $I_2^{(m)}$, we evaluate the original integral definitions by numerical quadrature at the same $(x_{mc},r)$ pairs and compare against the corresponding analytical values. Applying the same trigonometric substitution introduced in Section~\ref{sec:trig_sub}, $I_1^{(m)}$ and $I_2^{(m)}$ are evaluated by quadrature on $\phi\in[0,\phi_r]$ and $\phi\in[\phi_r,\pi/2]$, respectively, with smooth $\phi$-space integrands. Table~\ref{tab:validation_quadrature} confirms agreement between analytical and quadrature evaluations at the level of $10^{-17}$--$10^{-18}$, providing stringent end-to-end verification of the moment integral derivations.

\begin{table}[hbt!]
\caption{\label{tab:validation_quadrature} Independent quadrature verification of moment integrals.}
\centering
\begin{tabular}{lccccc}
\hline
Case & $x_{mc}$ & $r$ &
$|I_{1,\mathrm{anal}}^{(m)}-I_{1,\mathrm{quad}}^{(m)}|$ &
$|I_{2,\mathrm{anal}}^{(m)}-I_{2,\mathrm{quad}}^{(m)}|$ &
$|(I_1^{(m)}+\frac{1}{(1-r)^3}I_2^{(m)})_{\mathrm{quad}}|$ \\
\hline
221 & 0.10 & 0.13074976 & $2.17\times 10^{-17}$ & $2.44\times 10^{-18}$ & $1.04\times 10^{-17}$ \\
231 & 0.15 & 0.21601450 & $3.51\times 10^{-17}$ & $4.34\times 10^{-19}$ & $2.81\times 10^{-15}$ \\
241 & 0.20 & 0.31791890 & $3.82\times 10^{-17}$ & $3.04\times 10^{-18}$ & $1.05\times 10^{-15}$ \\
251 & 0.25 & 0.44083034 & $1.56\times 10^{-17}$ & $2.43\times 10^{-17}$ & $8.33\times 10^{-17}$ \\
\hline
\end{tabular}
\end{table}

\subsection{Independent Quadrature Verification of Lift Integrals}
\label{subsec:validation_quadrature_lift}

The same verification procedure was applied to the lift integrals $I_1^{(l)}$ and $I_2^{(l)}$. Table~\ref{tab:validation_quadrature_lift} confirms agreement between analytical and quadrature evaluations at the level of $10^{-17}$--$10^{-18}$, providing complete verification of the lift integral derivations.

\begin{table}[hbt!]
\caption{\label{tab:validation_quadrature_lift} Independent quadrature verification of lift integrals.}
\centering
\begin{tabular}{lcccc}
\hline
Case & $x_{mc}$ & $r$ &
$|I_{1,\mathrm{anal}}^{(l)}-I_{1,\mathrm{quad}}^{(l)}|$ &
$|I_{2,\mathrm{anal}}^{(l)}-I_{2,\mathrm{quad}}^{(l)}|$ \\
\hline
221 & 0.10 & 0.13074976 & $6.07\times 10^{-18}$ & $2.30\times 10^{-18}$ \\
231 & 0.15 & 0.21601450 & $2.78\times 10^{-17}$ & $4.55\times 10^{-18}$ \\
241 & 0.20 & 0.31791890 & $1.39\times 10^{-17}$ & $4.34\times 10^{-18}$ \\
251 & 0.25 & 0.44083034 & $0$ & $1.60\times 10^{-17}$ \\
\hline
\end{tabular}
\end{table}

\subsection{Validation of Scaling Constant \texorpdfstring{$k_1$}{k1}}
\label{subsec:validation_k1}

The scaling constant $k_1$ was computed using Eq.~\eqref{eq:k1_closed_form} with the analytically determined $r$ values and compared against the tabulated reference values~\cite{Ladson_Brooks_1975}. Table~\ref{tab:validation_k1} presents the results.

\begin{table}[hbt!]
\caption{\label{tab:validation_k1} Comparison of computed $k_1$ values against tabulations~\cite{Ladson_Brooks_1975}.}
\centering
\begin{tabular}{lcccccc}
\hline
Designation & $x_{mc}$ & $L$ & $(k_1)_{\mathrm{ref}}$ & $(k_1)_{\mathrm{anal}}$ & Absolute Error & Relative Error \\
\hline
221 & 0.10 & 2 & 51.990 & 51.120 & 0.870 & 1.67\% \\
231 & 0.15 & 2 & 15.793 & 15.691 & 0.102 & 0.65\% \\
241 & 0.20 & 2 & 6.520 & 6.507 & 0.013 & 0.19\% \\
251 & 0.25 & 2 & 3.191 & 3.176 & 0.015 & 0.48\% \\
\hline
\end{tabular}
\end{table}

Across these cases, $k_1$ is reproduced to within $1.7\%$ of the tabulations. The largest discrepancies occur for the 221 and 231 cases, where $k_1$ is large because the effective camber is small when $r \approx x_{mc}$. However, as demonstrated in Section~\ref{subsec:sensitivity}, these discrepancies cannot be attributed to the propagation of errors in $r$. A first-order sensitivity analysis predicts that the $\mathcal{O}(10^{-3})$ differences in $r$ would produce changes in $k_1$ of only $\mathcal{O}(10^{-2})$, whereas the observed discrepancies are $\mathcal{O}(10^{-1})$ to $\mathcal{O}(1)$.

The discrepancies, therefore, reflect differences in the original tabulation methodology rather than errors in the present derivation. Possible sources include different numerical quadrature schemes or limited floating-point precision in the legacy NACA-era computations, rounding of intermediate parameters before computing $k_1$, or minor differences in conventions or governing equations. Crucially, the closed-form lift integrals underlying Eq.~\eqref{eq:k1_closed_form} are independently verified against numerical quadrature to machine precision (Section~\ref{subsec:validation_quadrature_lift}), confirming the analytical correctness of the present expressions. The closed-form $k_1$ values may therefore represent a more accurate evaluation than the legacy tabulations.

\subsection{Verification of \texorpdfstring{$k_2/k_1$}{k2/k1} Ratio}
\label{subsec:validation_k2k1}

As a supplementary check, the ratio $k_2/k_1$ was computed using Eq.~\eqref{eq:k2k1_ratio} with the analytically determined $r$ values and then compared against the tabulated reference values~\cite{Ladson_Brooks_1975}. Table~\ref{tab:validation_k2k1} presents the results. Although the relative errors for the 221 and 231 cases appear large (up to $20\%$), this is a consequence of the small magnitude of $k_2/k_1$ rather than a deficiency in the analytical expressions. As confirmed by the sensitivity analysis in Section~\ref{subsec:sensitivity}, these discrepancies are fully explained by the propagation of $\mathcal{O}(10^{-3})$ differences in $r$: the discrepancies estimated from $\frac{\partial(k_2/k_1)}{\partial r} \cdot \Delta r$ match the observed values to within $\approx 10\%$ (Table~\ref{tab:sensitivity_k2k1}).

\begin{table}[hbt!]
\caption{\label{tab:validation_k2k1} Verification of $k_2/k_1$ ratio against tabulations~\cite{Ladson_Brooks_1975}.}
\centering
\begin{tabular}{lcccc}
\hline
Designation & $(k_2/k_1)_{\mathrm{ref}}$& $(k_2/k_1)_{\mathrm{anal}}$ & Absolute Error & Relative Error \\
\hline
221 & 0.000764 & 0.000916 & $1.52\times 10^{-4}$ & 19.85\% \\
231 & 0.006770 & 0.006213 & $5.57\times 10^{-4}$ & 8.22\% \\
241 & 0.030300 & 0.030195 & $1.05\times 10^{-4}$ & 0.35\% \\
251 & 0.135500 & 0.134878 & $6.22\times 10^{-4}$ & 0.46\% \\
\hline
\end{tabular}
\end{table}

From Eq.~\eqref{eq:k2k1_ratio}, the ratio $k_2/k_1$ depends on the numerator $3(r - x_{mc})^2 - r^3$, which approaches zero as $r \to x_{mc}$. Physically, a small $k_2/k_1$ indicates minimal reflex in the aft region, and in this limit the camber line is nearly cubic throughout; the ratio therefore becomes ill-conditioned with respect to $r$. For the 221 case, $(r - x_{mc}) \approx 0.03$ and the numerator is $\mathcal{O}(10^{-3})$. A discrepancy of $\mathcal{O}(10^{-4})$ in $r$ thus propagates to a comparable absolute error in $k_2/k_1$, but represents a large fraction of the small ratio. For the 241 and 251 cases, the breakpoint lies further aft, $(r - x_{mc})$ is larger, and $k_2/k_1$ is correspondingly of order $10^{-2}$ to $10^{-1}$; the same absolute errors in $r$ now produce relative errors below $0.5\%$. This confirms that the analytical expression is accurate and that the apparent discrepancies for forward-camber cases are a quantitatively predictable consequence of parametric sensitivity rather than analytical error.

\subsection{Sensitivity Analysis}
\label{subsec:sensitivity}
The discrepancies observed in Sections~\ref{subsec:validation_k1} and~\ref{subsec:validation_k2k1} can be examined through a sensitivity analysis measuring how small perturbations in $r$ propagate into $k_2/k_1$ and $k_1$.
\subsubsection{Sensitivity of \texorpdfstring{$k_2/k_1$}{k2/k1}}
From Eq.~\eqref{eq:k2k1_ratio}, the derivative of $k_2/k_1$ with respect to $r$ is obtained via the quotient rule:
\begin{equation}
\frac{\partial}{\partial r}\left(\frac{k_2}{k_1}\right) = \frac{\left[6(r - x_{mc}) - 3r^2\right](1-r)^3 + 3(1-r)^2\left[3(r - x_{mc})^2 - r^3\right]}{(1-r)^6}
\label{eq:dk2k1_dr}
\end{equation}
Table~\ref{tab:sensitivity_k2k1} presents the computed sensitivities and demonstrates that the observed discrepancies are well-predicted by the first-order approximation $\Delta(k_2/k_1)_{\mathrm{anal}} \approx \frac{\partial(k_2/k_1)}{\partial r} \cdot \Delta r$.
\begin{table}[hbt!]
\caption{\label{tab:sensitivity_k2k1} Sensitivity analysis for $k_2/k_1$: comparison of analytical and reference discrepancies.}
\centering
\begin{tabular}{lccccc}
\hline
Designation & $r$ & $\dfrac{\partial(k_2/k_1)}{\partial r}$ & $\Delta r$ & $\Delta(k_2/k_1)_{\mathrm{anal}}$ & $\Delta(k_2/k_1)_{\mathrm{ref}}$ \\[4pt]
\hline
221 & 0.1307 & 0.206 & $+7.50\times 10^{-4}$ & $+1.54\times 10^{-4}$ & $+1.52\times 10^{-4}$ \\
231 & 0.2160 & 0.555 & $-9.85\times 10^{-4}$ & $-5.47\times 10^{-4}$ & $-5.57\times 10^{-4}$ \\
241 & 0.3179 & 1.407 & $-8.10\times 10^{-5}$ & $-1.14\times 10^{-4}$ & $-1.05\times 10^{-4}$ \\
251 & 0.4408 & 3.938 & $-1.70\times 10^{-4}$ & $-6.68\times 10^{-4}$ & $-6.22\times 10^{-4}$ \\
\hline
\end{tabular}
\end{table}
The close agreement between analytical and reference discrepancies confirms that the $k_2/k_1$ differences arise entirely from the propagation of small errors in $r$. For the 221 case, $\partial(k_2/k_1)/\partial r \approx 0.21$. Although this sensitivity is modest, the ratio $k_2/k_1 \approx 0.0009$ is itself small, so even minor absolute changes appear as large relative errors. The sensitivity increases as the breakpoint moves aft, reaching $\partial(k_2/k_1)/\partial r \approx 3.9$ for the 251 case. Yet, the larger baseline value of $k_2/k_1 \approx 0.135$ renders these variations less significant in relative terms.
\subsubsection{Sensitivity of \texorpdfstring{$k_1$}{k1}}
The sensitivity of $k_1$ to perturbations in $r$ is computed numerically from Eq.~\eqref{eq:k1_closed_form}. Table~\ref{tab:sensitivity_k1} presents the results.
\begin{table}[hbt!]
\caption{\label{tab:sensitivity_k1} Sensitivity analysis for $k_1$: comparison of analytical and reference discrepancies.}
\centering
\begin{tabular}{lccccc}
\hline
Designation & $r$ & $\dfrac{\partial k_1}{\partial r}$ & $\Delta r$ & $\Delta k_{1,\mathrm{anal}}$ & $\Delta k_{1,\mathrm{ref}}$ \\[4pt]
\hline
221 & 0.1307 & $-13.5$ & $+7.50\times 10^{-4}$ & $-0.010$ & $-0.870$ \\
231 & 0.2160 & $+1.70$ & $-9.85\times 10^{-4}$ & $-0.002$ & $-0.102$ \\
241 & 0.3179 & $+2.52$ & $-8.10\times 10^{-5}$ & $-0.000$ & $-0.013$ \\
251 & 0.4408 & $+2.15$ & $-1.70\times 10^{-4}$ & $-0.000$ & $-0.015$ \\
\hline
\end{tabular}
\end{table}
Unlike $k_2/k_1$, the analytical discrepancies in $k_1$ from error propagation in $r$ do not explain the reference differences. For example, in the 221 case, the sensitivity $\partial k_1/\partial r \approx -13.5$ with $\Delta r \approx 7.5 \times 10^{-4}$ predicts $\Delta k_1 \approx -0.01$. However, the reference discrepancy is $-0.87$, nearly two orders of magnitude larger. This indicates that the $k_1$ discrepancies arise primarily from differences in how the original tabulations were computed, rather than propagation of $r$ errors alone. Possible sources include different numerical quadrature schemes or precision in legacy computations, intermediate value rounding before computing $k_1$, or slight differences in the governing equations or conventions used historically.

\subsection{Graphical Validation}
\label{subsec:graphical_validation}

\begin{figure}[hbt!]
\centering
\includegraphics[width=\textwidth]{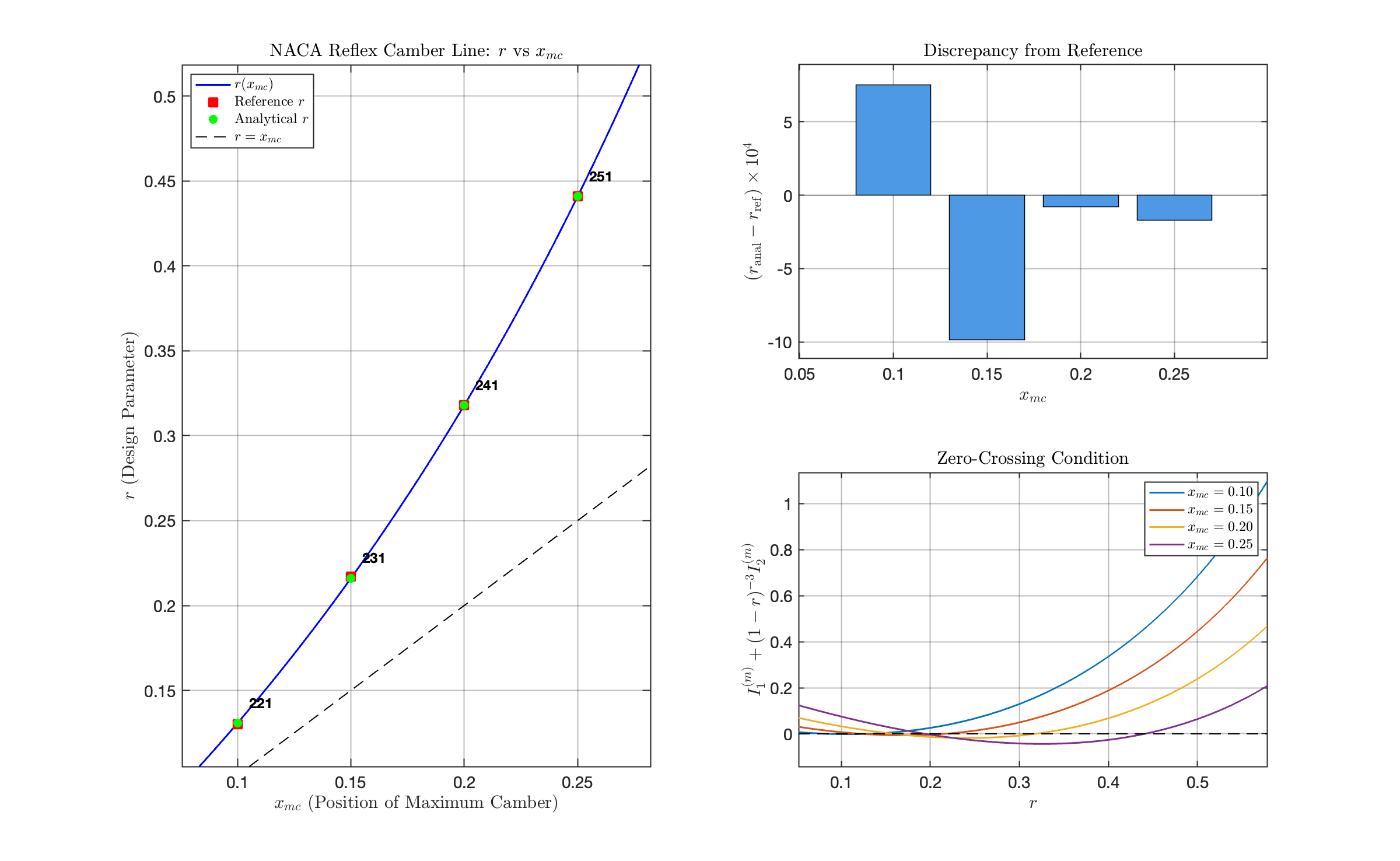}
\caption{Validation of breakpoint parameter $r$: (left) analytical and reference~\cite{Ladson_Brooks_1975} values versus $x_{mc}$; (upper right) discrepancy $r_{\mathrm{anal}}-r_{\mathrm{ref}}$; (lower right) moment-condition residual versus $r$.}
\label{fig:validation_r}
\end{figure}

Fig.~\ref{fig:validation_r} provides a visual summary of the agreement between the analytical $r(x_{mc})$ relation and the reference values~\cite{Ladson_Brooks_1975}. The left panel displays the analytical mapping of $r$ versus $x_{mc}$ (solid line), along with the reference points and analytical roots, which are visually coincident. The dashed line $r=x_{mc}$ indicates the boundary of the physically admissible region $r>x_{mc}$. The upper-right panel plots the signed discrepancy scaled by $10^4$, confirming deviations of order $10^{-3}$, while the lower-right panel shows $I_1^{(m)}+\frac{1}{(1-r)^3}I_2^{(m)}$ as a function of $r$ for each $x_{mc}$, demonstrating clear zero-crossings at the solution values.

\begin{figure}[hbt!]
\centering
\includegraphics[width=\textwidth]{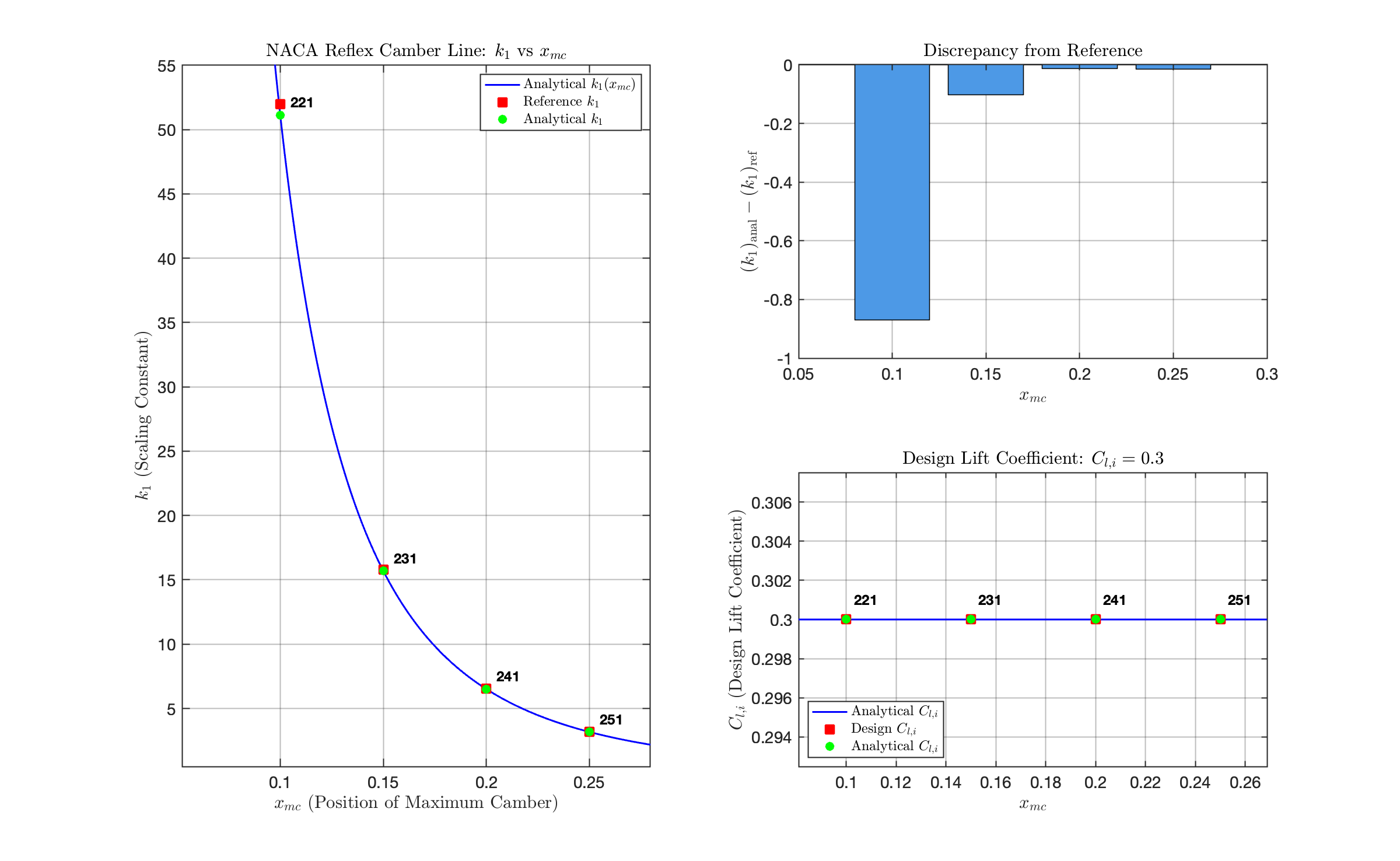}
\caption{Validation of scaling constant $k_1$: (left) analytical and reference~\cite{Ladson_Brooks_1975} values versus $x_{mc}$; (upper right) discrepancy $k_{1,\mathrm{anal}} - k_{1,\mathrm{ref}}$; (lower right) recovered $C_{l,i}$.}
\label{fig:validation_k1}
\end{figure}

Fig.~\ref{fig:validation_k1} presents corresponding results for the scaling constant $k_1$. The left panel shows the analytical $k_1(x_{mc})$ relation, which exhibits the expected asymptotic increase as $x_{mc}$ decreases (requiring larger $k_1$ to achieve the design lift when the effective camber is reduced). The upper-right panel displays the signed discrepancy $k_{1,\mathrm{anal}} - k_{1,\mathrm{ref}}$, showing the largest deviation for the 221 case. The lower-right panel verifies that the design lift coefficient $C_{l,i} = 0.3$ (corresponding to $L=2$) is correctly recovered across all cases, confirming internal consistency of the closed-form lift-integral formulation.

\subsection{Discussion}
\label{subsec:validation_discussion}

The validation results demonstrate that the closed-form analytical framework reproduces the reference design parameters~\cite{Ladson_Brooks_1975} to within $\mathcal{O}(10^{-3})$ for $r$ and $\mathcal{O}(10^{-2})$ for $k_1$ across the representative cases 221, 231, 241, and 251. Table~\ref{tab:validation_summary} summarises the principal validation metrics, and the sensitivity analysis (Section~\ref{subsec:sensitivity}) clarifies the origin and significance of the observed discrepancies.

\begin{table}[hbt!]
\caption{\label{tab:validation_summary} Summary of validation metrics (cases: 221, 231, 241, 251).}
\centering
\begin{tabular}{lc}
\hline
Metric & Value \\
\hline
Maximum $|r_{\mathrm{anal}} - r_{\mathrm{ref}}|$ & $9.85\times 10^{-4}$ \\
Mean $|r_{\mathrm{anal}} - r_{\mathrm{ref}}|$ & $4.97\times 10^{-4}$ \\
Maximum moment-condition residual $|I_1^{(m)}+\frac{1}{(1-r)^3}I_2^{(m)}|$ & $2.78\times 10^{-15}$ \\
Maximum $|k_{1,\mathrm{anal}} - k_{1,\mathrm{ref}}|$ & $0.87$ \\
Maximum relative error in $k_1$ & $1.67\%$ \\
Maximum $|(k_2/k_1)_{\mathrm{anal}} - (k_2/k_1)_{\mathrm{ref}}|$ & $6.22\times 10^{-4}$ \\
\hline
\end{tabular}
\end{table}

The analytical values of the breakpoint parameter $r$ satisfy the zero-moment condition to machine precision ($10^{-15}$--$10^{-17}$), and agree with the reference tabulations to within the expected rounding uncertainty associated with four-decimal-place reporting. In this sense, the closed-form moment framework should be interpreted as providing higher-precision breakpoint locations consistent with the original condition, rather than merely reproducing rounded reference values.

The discrepancies in the ratio $k_2/k_1$ between analytical and reference values are quantitatively explained by the propagation of $\mathcal{O}(10^{-3})$ differences in $r$ through the sensitivity $\partial(k_2/k_1)/\partial r$. This behaviour is consistent with the first-order estimate
\begin{equation}
    \Delta\!\left(\frac{k_2}{k_1}\right)\approx
    \frac{\partial(k_2/k_1)}{\partial r}\,\Delta r
\end{equation}
and confirms that the closed-form expression for $k_2/k_1$ is verified. Practically, this implies that the reference scatter in $k_2/k_1$ is an expected consequence of rounding in $r$, rather than an inconsistency in the present formulation.

In contrast, the observed discrepancies in the scaling constant $k_1$ are not explained by $r$ error propagation: the analytical perturbations based on $\partial k_1/\partial r$ are $\mathcal{O}(10^{-2})$, while the reference differences are $\mathcal{O}(10^{-1})$ to $\mathcal{O}(1)$ for the cases considered. Given that the lift-condition integrals are independently verified by numerical quadrature to machine precision (Table~\ref{tab:validation_quadrature_lift}), the most plausible explanation is that the reference $k_1$ values reflect differences in legacy computational methodology, such as historical quadrature, intermediate rounding or implementation conventions, rather than deficiencies in the present closed-form derivation.

The sensitivity results further show that when $r \approx x_{mc}$, a regime characteristic of forward-camber reflex configurations, both $k_2/k_1$ and $k_1$ become more sensitive to the breakpoint location. In this regime, small perturbations in $r$ can induce disproportionately large changes in the derived constants, thus underscoring the value of computing $r$ to higher precision from the exact zero-moment condition rather than relying on four-decimal-place tabulations.

Finally, the analytical expressions are corroborated independently by direct numerical quadrature of the original integral definitions, yielding agreement at the level of $10^{-17}$--$10^{-18}$ for the moment integrals (Table~\ref{tab:validation_quadrature}) and $10^{-17}$--$10^{-18}$ for the lift integrals (Table~\ref{tab:validation_quadrature_lift}). This provides stringent verification of the closed-form results and supports the use of the present framework as a higher-precision alternative to reference tabulations.

\section{Extended Design Parameters}\label{sec:extended_parameters}

The closed-form expressions enable computation of design parameters for any valid NACA 5-digit reflex designation. Table~\ref{tab:extended_parameters} presents values for the standard ranges $P \in [1,5]$ and $L \in [1,6]$, corresponding to $x_{mc} \in [0.05, 0.25]$ and $C_{l,i} \in [0.15, 0.90]$. Beyond $P = 5$, the ratio $k_2/k_1$ increases rapidly, indicating impractically large reflex curvature. Consistently, for $P = 9$ Eq.~\eqref{eq:r_transcendental} admits no admissible root $r\in(x_{mc},1)$. The tabulated values of $r$ and $k_2/k_1$ depend only on $x_{mc}$ and are therefore constant within each group of fixed $P$, while $k_1$ and $k_2$ scale linearly with the design lift coefficient; the latter is computed simply as $k_2 = (k_2/k_1) \cdot k_1$.

\begin{table}[hbt!]
\caption{\label{tab:extended_parameters} Design parameters for NACA 5-digit reflex camber lines.}
\centering
\begin{tabular}{cccccc}
\hline
Camber-line & & & & & \\
designation & $x_{mc}$ & $r$ & $k_1$ & $k_2/k_1$ & $k_2$ \\
\hline
111 & 0.05 & 0.0591 & 174.582 & 0.000051 & 0.008904 \\
121 & 0.10 & 0.1307 & 25.560 & 0.000916 & 0.023413 \\
131 & 0.15 & 0.2160 & 7.845 & 0.006213 & 0.048742 \\
141 & 0.20 & 0.3179 & 3.254 & 0.030195 & 0.098255 \\
151 & 0.25 & 0.4408 & 1.588 & 0.134878 & 0.214186 \\
\hline
211 & 0.05 & 0.0591 & 349.163 & 0.000051 & 0.017807 \\
221 & 0.10 & 0.1307 & 51.120 & 0.000916 & 0.046826 \\
231 & 0.15 & 0.2160 & 15.691 & 0.006213 & 0.097488 \\
241 & 0.20 & 0.3179 & 6.507 & 0.030195 & 0.196496 \\
251 & 0.25 & 0.4408 & 3.176 & 0.134878 & 0.428332 \\
\hline
311 & 0.05 & 0.0591 & 523.745 & 0.000051 & 0.026711 \\
321 & 0.10 & 0.1307 & 76.680 & 0.000916 & 0.070239 \\
331 & 0.15 & 0.2160 & 23.536 & 0.006213 & 0.146230 \\
341 & 0.20 & 0.3179 & 9.761 & 0.030195 & 0.294751 \\
351 & 0.25 & 0.4408 & 4.763 & 0.134878 & 0.642439 \\
\hline
411 & 0.05 & 0.0591 & 698.327 & 0.000051 & 0.035615 \\
421 & 0.10 & 0.1307 & 102.240 & 0.000916 & 0.093652 \\
431 & 0.15 & 0.2160 & 31.382 & 0.006213 & 0.194976 \\
441 & 0.20 & 0.3179 & 13.015 & 0.030195 & 0.392993 \\
451 & 0.25 & 0.4408 & 6.351 & 0.134878 & 0.856624 \\
\hline
511 & 0.05 & 0.0591 & 872.908 & 0.000051 & 0.044518 \\
521 & 0.10 & 0.1307 & 127.801 & 0.000916 & 0.117066 \\
531 & 0.15 & 0.2160 & 39.227 & 0.006213 & 0.243718 \\
541 & 0.20 & 0.3179 & 16.268 & 0.030195 & 0.491234 \\
551 & 0.25 & 0.4408 & 7.939 & 0.134878 & 1.070810 \\
\hline
611 & 0.05 & 0.0591 & 1047.490 & 0.000051 & 0.053422 \\
621 & 0.10 & 0.1307 & 153.361 & 0.000916 & 0.140479 \\
631 & 0.15 & 0.2160 & 47.073 & 0.006213 & 0.292464 \\
641 & 0.20 & 0.3179 & 19.522 & 0.030195 & 0.589489 \\
651 & 0.25 & 0.4408 & 9.527 & 0.134878 & 1.284996 \\
\hline
\end{tabular}
\end{table}

\section{Conclusion}
\label{sec:conclusion}
Closed-form analytical expressions have been derived for all thin-airfoil integrals arising in the design of NACA 5-digit reflex camber lines. The moment integrals $I_1^{(m)}$ and $I_2^{(m)}$ admit closed-form representations involving $\arcsin(\sqrt{r})$, $\arccos(\sqrt{r})$, and polynomial contributions $\mathcal{P}$, $\mathcal{Q}$, $\mathcal{R}$ in $r$ and $x_{mc}$, while the lift integrals $I_1^{(l)}$ and $I_2^{(l)}$ admit analogous representations with polynomial contributions $\mathcal{S}$, $\mathcal{T}$, $\mathcal{U}$. The breakpoint parameter $r$ is obtained by solving the transcendental condition $I_1^{(m)}+\frac{1}{(1-r)^3}I_2^{(m)}=0$ using one-dimensional root-finding, thus requiring no numerical integration; the resulting values satisfy the zero-moment condition to machine precision and provide higher-precision breakpoint locations than four-decimal-place legacy tabulations. The scaling constant $k_1$ follows from direct closed-form evaluation, rendering the complete design workflow quadrature-free. The observed differences in $k_2/k_1$ relative to legacy tabulations are explained quantitatively by propagation of rounding errors in $r$, constituting indirect verification of the closed-form expressions; in regimes where $r \approx x_{mc}$, the elevated sensitivity of both $k_2/k_1$ and $k_1$ makes the higher-precision $r$ values particularly valuable.

The resulting formulation replaces repeated numerical quadrature with analytically evaluable expressions, therefore improving reproducibility and computational efficiency. By eliminating numerical integration from the design process, the present work removes a source of variability that has persisted in the literature for nearly a century. Independent quadrature verification of both moment and lift integrals to machine precision confirms the analytical correctness of the present expressions, which may therefore represent a more accurate evaluation than legacy tabulations. The extended parameter tables provided herein offer a comprehensive reference for designations beyond the limited set available in historical NACA tabulations. The methodology presented here may be extended to other piecewise-defined camber-line families or adapted for use in automated airfoil optimization frameworks; these closed-form expressions may prove particularly useful in gradient-based optimization, where analytical derivatives of the design parameters with respect to the input variables can now be obtained directly.

\appendix

\section{Derivation of Moment Integrals}\label{sec:moment_derivation}

\subsection{\texorpdfstring{$I_1^{(m)}$}{I1m}: Region \texorpdfstring{$0 \leq x < r$}{0 <= x < r}}\label{sec:region_1}

From Eq.~\eqref{eq:I1m_def},
\begin{equation}
I_1^{(m)} = \int_0^{r} \left(x^2-2rx+2x_{mc}r-x_{mc}^2\right)\frac{8x^2-6x}{\sqrt{x(1-x)}}\,\mathrm{d}x
\end{equation}

Expanding the polynomial product:
\begin{equation}
\left(x^2-2rx+2x_{mc}r-x_{mc}^2\right)(8x^2-6x) = a_4 x^4 + a_3 x^3 + a_2 x^2 + a_1 x
\end{equation}
with
\begin{align}
a_4 &= 8 \\
a_3 &= -16r-6 \\
a_2 &= 12r+16x_{mc}r-8x_{mc}^2 \\
a_1 &= 6x_{mc}^2-12x_{mc}r
\end{align}

Applying the substitution $x=\sin^2\phi$ with $\mathrm{d}x/\sqrt{x(1-x)}=2\,\mathrm{d}\phi$:
\begin{equation}
I_1^{(m)} = 2\int_0^{\phi_r} \left[a_4\sin^8\phi + a_3\sin^6\phi + a_2\sin^4\phi + a_1\sin^2\phi\right] \mathrm{d}\phi
\end{equation}
where $\phi_r \equiv \arcsin(\sqrt{r})$.

Using the reduction formulas:
\begin{align}
\int \sin^2\phi \, \mathrm{d}\phi &= \frac{\phi}{2} - \frac{\sin\phi\cos\phi}{2} \\
\int \sin^4\phi \, \mathrm{d}\phi &= \frac{3\phi}{8} - \frac{3\sin\phi\cos\phi}{8} - \frac{\sin^3\phi\cos\phi}{4} \\
\int \sin^6\phi \, \mathrm{d}\phi &= \frac{5\phi}{16} - \frac{5\sin\phi\cos\phi}{16} - \frac{5\sin^3\phi\cos\phi}{24} - \frac{\sin^5\phi\cos\phi}{6} \\
\int \sin^8\phi \, \mathrm{d}\phi &= \frac{35\phi}{128} - \frac{35\sin\phi\cos\phi}{128} - \frac{35\sin^3\phi\cos\phi}{192} - \frac{7\sin^5\phi\cos\phi}{48} - \frac{\sin^7\phi\cos\phi}{8}
\end{align}

Substituting and grouping:
\begin{equation}
I_1^{(m)} = 2\left[A\,\phi + B\,\sin\phi\cos\phi + C_3\,\sin^3\phi\cos\phi + C_5\,\sin^5\phi\cos\phi + C_7\,\sin^7\phi\cos\phi\right]_0^{\phi_r}
\end{equation}
where
\begin{align}
A &= \frac{35a_4}{128} + \frac{5a_3}{16} + \frac{3a_2}{8} + \frac{a_1}{2} \\
B &= -A \\
C_3 &= -\frac{35a_4}{192} - \frac{5a_3}{24} - \frac{a_2}{4} \\
C_5 &= -\frac{7a_4}{48} - \frac{a_3}{6} \\
C_7 &= -\frac{a_4}{8}
\end{align}

At $\phi=0$ all terms vanish. At $\phi=\phi_r$, using $\sin\phi_r=\sqrt{r}$ and $\cos\phi_r=\sqrt{1-r}$:
\begin{equation}
I_1^{(m)} = 2A\,\arcsin(\sqrt{r}) + 2\sqrt{r(1-r)}\left(B + C_3 r + C_5 r^2 + C_7 r^3\right)
\end{equation}

Substituting $a_4, a_3, a_2, a_1$ and simplifying:
\begin{equation}
2A = \frac{5-8r}{8}
\end{equation}
\begin{equation}
2\left(B + C_3 r + C_5 r^2 + C_7 r^3\right) = \mathcal{P}(r, x_{mc})
\end{equation}
where
\begin{equation}
\mathcal{P}(r,x_{mc}) = 4r(x_{mc}-r)^2 + \frac{-16r^3+8r^2+14r-15}{24}
\end{equation}

Therefore:
\begin{equation}
I_1^{(m)} = \frac{5-8r}{8}\,\arcsin(\sqrt{r}) + \sqrt{r(1-r)}\,\mathcal{P}(r,x_{mc})
\end{equation}

\subsection{\texorpdfstring{$I_2^{(m)}$}{I2m}: Region \texorpdfstring{$r \leq x \leq 1$}{r <= x <= 1}}\label{sec:region_2}

From Eq.~\eqref{eq:I2m_def},
\begin{equation}
I_2^{(m)} = \int_r^{1} \left(b_2x^2+b_1x+b_0\right)\frac{8x^2-6x}{\sqrt{x(1-x)}}\,\mathrm{d}x
\end{equation}
with $b_2$, $b_1$, $b_0$ as defined in Eqs.~\eqref{eq:b2}--\eqref{eq:b0}.

Expanding the polynomial product:
\begin{equation}
\left(b_2 x^2 + b_1 x + b_0\right)(8x^2-6x) = c_4 x^4 + c_3 x^3 + c_2 x^2 + c_1 x
\end{equation}
with
\begin{align}
c_4 &= 8b_2 \\
c_3 &= -6b_2 + 8b_1 \\
c_2 &= -6b_1 + 8b_0 \\
c_1 &= -6b_0
\end{align}

Applying the substitution $x=\sin^2\phi$:
\begin{equation}
I_2^{(m)} = 2\int_{\phi_r}^{\pi/2} \left[c_4\sin^8\phi + c_3\sin^6\phi + c_2\sin^4\phi + c_1\sin^2\phi\right] \mathrm{d}\phi
\end{equation}

Using the same reduction formulas and grouping:
\begin{equation}
I_2^{(m)} = 2\left[A'\,\phi + B'\,\sin\phi\cos\phi + C'_3\,\sin^3\phi\cos\phi + C'_5\,\sin^5\phi\cos\phi + C'_7\,\sin^7\phi\cos\phi\right]_{\phi_r}^{\pi/2}
\end{equation}
where $A'$, $B'$, $C'_3$, $C'_5$, $C'_7$ follow the same structure as before with $c_i$ replacing $a_i$.

At $\phi=\pi/2$, all $\sin^{2n+1}\phi\cos\phi$ terms vanish. Using $\frac{\pi}{2}-\phi_r=\arccos(\sqrt{r})$:
\begin{equation}
I_2^{(m)} = 2A'\,\arccos(\sqrt{r}) - 2\sqrt{r(1-r)}\left(B' + C'_3 r + C'_5 r^2 + C'_7 r^3\right)
\end{equation}

Substituting and simplifying:
\begin{equation}
2A' = \mathcal{Q}(r, x_{mc})
\end{equation}
\begin{equation}
2\left(B' + C'_3 r + C'_5 r^2 + C'_7 r^3\right) = \mathcal{R}(r, x_{mc})
\end{equation}
where
\begin{equation}
\mathcal{Q}(r,x_{mc}) = \frac{8r-5}{8}\left[r^3-3(x_{mc}-r)^2\right]
\end{equation}
\begin{equation}
\mathcal{R}(r,x_{mc}) = \frac{r^3}{24}(16r^3-8r^2-14r+15) - \frac{1}{8}(32r^4-80r^3+88r^2-46r+15)(x_{mc}-r)^2
\end{equation}

Therefore:
\begin{equation}
I_2^{(m)} = \mathcal{Q}(r,x_{mc})\,\arccos(\sqrt{r}) - \sqrt{r(1-r)}\,\mathcal{R}(r,x_{mc})
\end{equation}

\section{Derivation of Lift Integrals}\label{app:lift_derivation}

\subsection{\texorpdfstring{$I_1^{(l)}$}{I1l}: Region \texorpdfstring{$0 \leq x < r$}{0 <= x < r}}\label{sec:lift_region_1}

From Eq.~\eqref{eq:I1l_def},
\begin{equation}
I_1^{(l)} = \int_0^{r} \left(x^2-2rx+2x_{mc}r-x_{mc}^2\right)\frac{1-2x}{\sqrt{x(1-x)}}\,\mathrm{d}x
\end{equation}

Expanding the polynomial product:
\begin{equation}
\left(x^2-2rx+2x_{mc}r-x_{mc}^2\right)(1-2x) = a_3 x^3 + a_2 x^2 + a_1 x + a_0
\end{equation}
with
\begin{align}
a_3 &= -2 \\
a_2 &= 1+4r \\
a_1 &= -2r-4x_{mc}r+2x_{mc}^2 \\
a_0 &= 2x_{mc}r-x_{mc}^2
\end{align}

Applying the substitution $x=\sin^2\phi$ with $\mathrm{d}x/\sqrt{x(1-x)}=2\,\mathrm{d}\phi$:
\begin{equation}
I_1^{(l)} = 2\int_0^{\phi_r} \left[a_3\sin^6\phi + a_2\sin^4\phi + a_1\sin^2\phi + a_0\right] \mathrm{d}\phi
\end{equation}
where $\phi_r \equiv \arcsin(\sqrt{r})$.

Using the reduction formulas:
\begin{align}
\int 1 \, \mathrm{d}\phi &= \phi \\
\int \sin^2\phi \, \mathrm{d}\phi &= \frac{\phi}{2} - \frac{\sin\phi\cos\phi}{2} \\
\int \sin^4\phi \, \mathrm{d}\phi &= \frac{3\phi}{8} - \frac{3\sin\phi\cos\phi}{8} - \frac{\sin^3\phi\cos\phi}{4} \\
\int \sin^6\phi \, \mathrm{d}\phi &= \frac{5\phi}{16} - \frac{5\sin\phi\cos\phi}{16} - \frac{5\sin^3\phi\cos\phi}{24} - \frac{\sin^5\phi\cos\phi}{6}
\end{align}

Substituting and grouping:
\begin{equation}
I_1^{(l)} = 2\left[A\,\phi + B\,\sin\phi\cos\phi + C_3\,\sin^3\phi\cos\phi + C_5\,\sin^5\phi\cos\phi\right]_0^{\phi_r}
\end{equation}
where
\begin{align}
A &= \frac{5a_3}{16} + \frac{3a_2}{8} + \frac{a_1}{2} + a_0 \\
B &= -\frac{5a_3}{16} - \frac{3a_2}{8} - \frac{a_1}{2} \\
C_3 &= -\frac{5a_3}{24} - \frac{a_2}{4} \\
C_5 &= -\frac{a_3}{6}
\end{align}

At $\phi=0$ all terms vanish. At $\phi=\phi_r$, using $\sin\phi_r=\sqrt{r}$ and $\cos\phi_r=\sqrt{1-r}$:
\begin{equation}
I_1^{(l)} = 2A\,\arcsin(\sqrt{r}) + 2\sqrt{r(1-r)}\left(B + C_3 r + C_5 r^2\right)
\end{equation}

Substituting $a_3, a_2, a_1, a_0$ and simplifying:
\begin{equation}
2A = r - \frac{1}{2}
\end{equation}
\begin{equation}
2\left(B + C_3 r + C_5 r^2\right) = \mathcal{S}(r, x_{mc})
\end{equation}
where
\begin{equation}
\mathcal{S}(r,x_{mc}) = \frac{1}{3}+\frac{2}{3}\left(r-\frac{1}{2}\right)^2-2(x_{mc}-r)^2
\end{equation}

Therefore:
\begin{equation}
I_1^{(l)} = \left(r-\frac{1}{2}\right)\arcsin(\sqrt{r}) + \sqrt{r(1-r)}\,\mathcal{S}(r,x_{mc})
\end{equation}

\subsection{\texorpdfstring{$I_2^{(l)}$}{I2l}: Region \texorpdfstring{$r \leq x \leq 1$}{r <= x <= 1}}\label{sec:lift_region_2}

From Eq.~\eqref{eq:I2l_def},
\begin{equation}
I_2^{(l)} = \int_r^{1} \left(b_2x^2+b_1x+b_0\right)\frac{1-2x}{\sqrt{x(1-x)}}\,\mathrm{d}x
\end{equation}
with $b_2$, $b_1$, $b_0$ as defined in Eqs.~\eqref{eq:b2}--\eqref{eq:b0}.

Expanding the polynomial product:
\begin{equation}
\left(b_2 x^2 + b_1 x + b_0\right)(1-2x) = c_3 x^3 + c_2 x^2 + c_1 x + c_0
\end{equation}
with
\begin{align}
c_3 &= -2b_2 \\
c_2 &= b_2 - 2b_1 \\
c_1 &= b_1 - 2b_0 \\
c_0 &= b_0
\end{align}

Applying the substitution $x=\sin^2\phi$:
\begin{equation}
I_2^{(l)} = 2\int_{\phi_r}^{\pi/2} \left[c_3\sin^6\phi + c_2\sin^4\phi + c_1\sin^2\phi + c_0\right] \mathrm{d}\phi
\end{equation}

Using the same reduction formulas and grouping:
\begin{equation}
I_2^{(l)} = 2\left[A'\,\phi + B'\,\sin\phi\cos\phi + C'_3\,\sin^3\phi\cos\phi + C'_5\,\sin^5\phi\cos\phi\right]_{\phi_r}^{\pi/2}
\end{equation}
where $A'$, $B'$, $C'_3$, $C'_5$ follow the same structure as before with $c_i$ replacing $a_i$.

At $\phi=\pi/2$, all $\sin^{2n+1}\phi\cos\phi$ terms vanish. Using $\frac{\pi}{2}-\phi_r=\arccos(\sqrt{r})$:
\begin{equation}
I_2^{(l)} = 2A'\,\arccos(\sqrt{r}) - 2\sqrt{r(1-r)}\left(B' + C'_3 r + C'_5 r^2\right)
\end{equation}

Substituting and simplifying:
\begin{equation}
2A' = \mathcal{T}(r, x_{mc})
\end{equation}
\begin{equation}
2\left(B' + C'_3 r + C'_5 r^2\right) = \mathcal{U}(r, x_{mc})
\end{equation}
where
\begin{equation}
\mathcal{T}(r,x_{mc}) = \left(\frac{1}{2}-r\right)\left[r^3-3(x_{mc}-r)^2\right]
\end{equation}
\begin{equation}
\mathcal{U}(r,x_{mc}) = \frac{1}{2}(4r^3-8r^2+8r-1)(x_{mc}-r)^2 - \frac{r^3}{6}(4r^2-4r+3)
\end{equation}

Therefore:
\begin{equation}
I_2^{(l)} = \mathcal{T}(r,x_{mc})\,\arccos(\sqrt{r}) - \sqrt{r(1-r)}\,\mathcal{U}(r,x_{mc})
\end{equation}

\section{Extended-Precision Breakpoint Values}\label{app:validation_precision}

For reproducibility, Table~\ref{tab:validation_extended} reports the analytical breakpoint parameter $r$ to eight decimal places for the cases considered in Section~\ref{subsec:validation_r}. These values satisfy the zero-moment condition to machine precision.

\begin{table}[hbt!]
\caption{\label{tab:validation_extended} Extended-precision breakpoint parameter $r$ for representative reflex configurations.}
\centering
\begin{tabular}{lcccc}
\hline
Designation & $x_{mc}$ & $r_{\mathrm{ref}}$ & $r_{\mathrm{anal}}$ & $r_{\mathrm{anal}}-r_{\mathrm{ref}}$ \\
\hline
221 & 0.10 & 0.1300 & 0.13074976 & $+7.4976\times 10^{-4}$ \\
231 & 0.15 & 0.2170 & 0.21601450 & $-9.8550\times 10^{-4}$ \\
241 & 0.20 & 0.3180 & 0.31791890 & $-8.1100\times 10^{-5}$ \\
251 & 0.25 & 0.4410 & 0.44083034 & $-1.6966\times 10^{-4}$ \\
\hline
\end{tabular}
\end{table}

\section*{Funding Sources}

No external funding was received for this research.

\section*{Acknowledgments}

The author gratefully acknowledges the foundational work of Ira H. Abbott and Albert E. von Doenhoff of the National Advisory Committee for Aeronautics, and the historical airfoil-ordinate tabulations compiled by Charles L. Ladson and Cuyler W. Brooks Jr. of NASA Langley Research Center, which provided the essential reference data for validation. AI-assisted tools (Claude and ChatGPT) were used for grammar checking, code assistance, and manuscript formatting; the author assumes full responsibility for the technical content and has verified all results independently.

\bibliographystyle{new-aiaa}
\bibliography{references}

\end{document}